\documentclass[fleqn,twoside]{article}
\usepackage[headings]{espcrc2}
\usepackage{epsfig}

\readRCS
$Id: espcrc2.tex,v 1.2 2004/02/24 11:22:11 spepping Exp $
\usepackage{graphicx}
\usepackage[figuresright]{rotating}

\newcommand{\AmS}{{\protect\the\textfont2
  A\kern-.1667em\lower.5ex\hbox{M}\kern-.125emS}}

\title{Pion scattering lengths from NA48}

\author{D. Madigozhin\address[MCSD]{Joint Institute for Nuclear Research, 
141980, Dubna, Moscow region, Russia}
\thanks{On behalf of the NA48/2 Collaboration: 
Cambridge, CERN, Chicago, Dubna, Edinburgh, Ferrara, Firenze, Mainz, Northwestern,
Perugia, Pisa, Saclay, Siegen, Torino, Vienna}
}

\begin{document}

\begin{abstract}
The anomaly (cusp) in the mass 
spectrum of $2\pi^0$ from $K^{\pm} \rightarrow \pi^{\pm} \pi^0\pi^0$ decay 
has been observed in the NA48/2 experiment. Using the recently
developed interpretation of this effect in terms of
Chiral Perturbative Theory, the pion scattering parameters are
measured. The preliminary result for the pion scattering length difference, 
based on the part of statistics,  
is $(a_0-a_2)m_{\pi^+} = 0.281 \pm 0.007 (stat)  \pm 0.014(syst) \pm 0.014(extern)$. 

\vspace{1pc}
\end{abstract}

% typeset front matter (including abstract)
\maketitle

\section{Beam and detectors}

The NA48/2 experiment at the CERN SPS is searching for the charge asymmetry in the decays
$K^{\pm} \rightarrow \pi^{\pm}\pi^+\pi^-$ and  $K^{\pm} \rightarrow \pi^{\pm}\pi^0\pi^0$. 

The experiment uses two simultaneous oppositely charged beams with a central momentum
of 60 GeV/c and a momentum band $\pm 3.8 \%$ , propagating along the same beam line. 
The decay volume is a 114 m long vacuum tank with a diameter of 1.92 m for the first 66 m, and 
2.4 m for the rest. The detected particles are dominated by the $K^{\pm}$ decay products. 

Charged particles are measured by the magnetic spectrometer, consisting of four drift chambers 
\cite{chambers} and a dipole magnet, located between the second and third chambers. Charged particles
are magnetically deflected in the horizontal plane by an angle corresponding to momentum kick of
120 MeV/c. The momentum resolution of the spectrometer is $\sigma(p)/p = 1.02\% \oplus 0.044\% p$ ($p$ in
GeV/c). 

The magnetic spectrometer is followed by a scintillator hodoscope consisting of two planes segmented
into horizontal and vertical strips and arranged in four quadrants.

A liquid krypton calorimeter \cite{lkr93} is used to reconstruct $\pi^0 \rightarrow \gamma\gamma$
decays. It is segmented transversally into $2 \times 2$ cm cells by a system of Cu-Be electrodes.
The calorimeter is 27 $X_0$ thick and absorbs the electromagnetic showers almost completely. Its
energy resolution for gamma-quanta is $\sigma(E)/E = 0.032/\sqrt{E} \oplus 0.09/E \oplus 0.0042,$
E in GeV. The space resolution for single electromagnetic showers can be parametrized as 
$\sigma_x = \sigma_y = 0.42 / \sqrt{E} \oplus 0.06$ cm for each transverse coordinate $x,y$.

A neutral hodoscope consisting of a plane of scintillating fibres is installed in the
liquid krypton calorimeter at a depth $9.5 X_0$. It is divided into four quadrants, each
consisting of eight bundles of vertical fibres optically connected to photomultiplier 
tubes.

\section{Trigger and events selection}

Data have been collected in 2003 and 2004, providing samples
of about $10^{8}$ $K^{\pm} \rightarrow \pi^{\pm}\pi^0\pi^0$ decays. Here we report the
analysis results of a partial sample $\approx 2.3 \times 10^{7}$ decays recorded in 2003.   

$K^{\pm} \rightarrow \pi^{\pm} \pi^0\pi^0$ decays are selected by a two-level trigger. The first
level requires a signal in at least one quadrant of the scintillator hodoscope in coincidence 
with the presence of energy depositions in calorimeter consistent with at least two photons. At
the second level, a fast on-line processor reconstructs the momentum of charged tracks and
calculates the missing mass, assuming the decay $K^{\pm} \rightarrow \pi^{\pm} + X$ of a 60 GeV/c kaon 
flying along the beam axis. The requirement that the mass of missed system $X$ is not consistent with 
the $\pi^0$ rejects the main $K^{\pm} \rightarrow \pi^{\pm} \pi^0$ background.

Events with at least one charged particle having a momentum above 5 GeV/c, and at least
four energy deposit clusters in liquid krypton calorimeter, each above the energy threshold
of 3 GeV and close in time to the track, are selected for the further analysis. The distance
between any two photons in calorimeter is required to be larger than 10 cm, and the distance
between each photon and the impact point of any track on calorimeter must exceed 15 cm. Fiducial
cuts on the distance of each photon from the calorimeter edges and centre are applied to avoid
the shower energy leakage. The distance between the charged particle track and the beam axis at 
the first drift chamber is required to be larger than 12 cm.

For each possible pair of photons we assume that it originates from $\pi^0 \rightarrow \gamma \gamma$
decay, and calculate the distance between the decay vertex and the front plane of calorimeter:

\begin{equation}
D_{ik} = \frac{\sqrt{E_iE_k[(x_i-x_k)^2+(y_i-y_k)^2]}}{m_{\pi^0}}
\end{equation}

where $E_i$, $E_k$ are the energies of the two photons and $x,y$ are their impact
point coordinates on calorimeter.

The two photon pairs with the smallest $D_{ik}$ difference are selected as the best
combination consistent with two $\pi^0$ mesons from $K^{\pm} \rightarrow \pi^{\pm} \pi^0\pi^0$
decay, and the arithmetic average of two $D$ values is used as the decay vertex position.
The final event selection requires that the $\pi^{\pm} \pi^0\pi^0$ invariant mass differs 
from the PDG charged kaon mass by at most $\pm 6$ MeV.  

\section{Cusp effect}

\begin{figure}[!h]
\epsfig{file=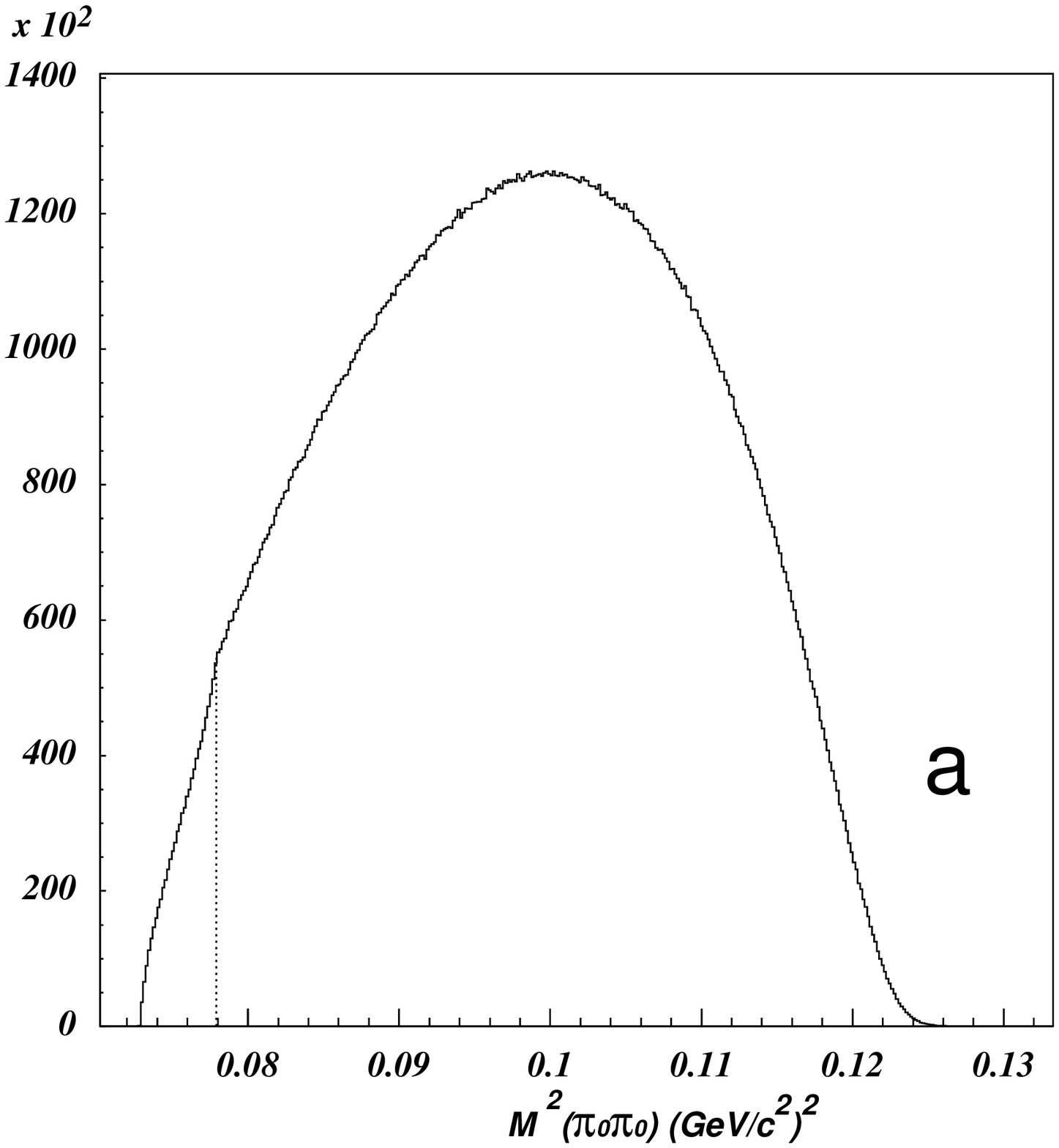,bbllx=16pt,bblly=140pt,bburx=567pt,bbury=700pt,width=68mm}
\epsfig{file=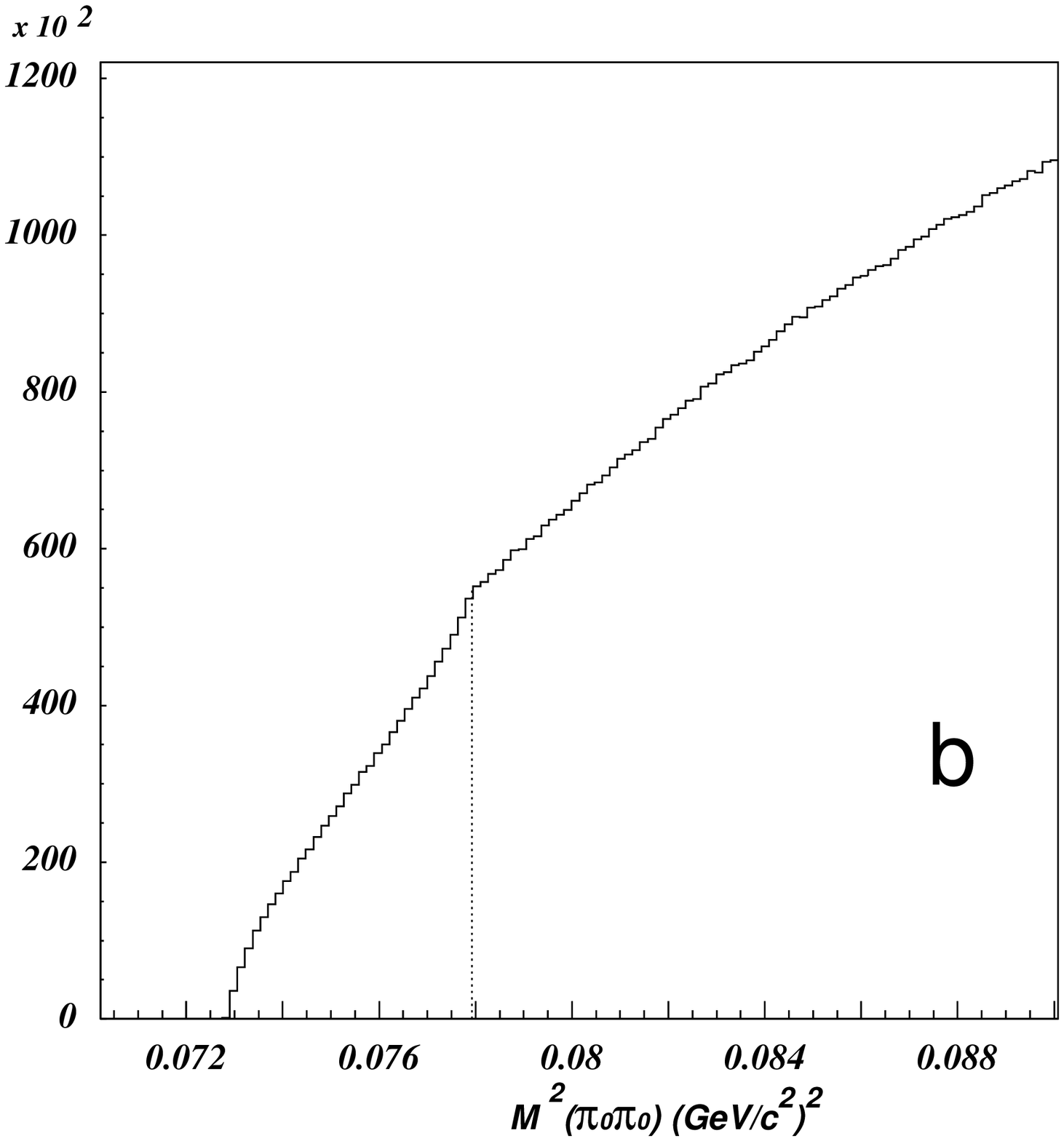,bbllx=16pt,bblly=140pt,bburx=567pt,bbury=700pt,width=68mm}
\caption{
The $M^2(\pi^0\pi^0)$ distribution for the subsystem of  the reconstructed 
$K^{\pm}\rightarrow \pi^{\pm}\pi^0\pi^0$ decays:
(a) -- in the full kinematic region; 
(b) -- in the region around the threshold;
the value corresponding to $4m^{2}_{\pi^{\pm}}$
is indicated by vertical line.
} 
\label{S2_1}
\end{figure}

The invariant mass of $\pi^0\pi^0$ subsystem ($M_{00}$) in $K^{\pm} \rightarrow \pi^{\pm} \pi^0 \pi^0$ decays 
has been studied to
search for possible anomaly at the threshold region: $M_{00} = 2\cdot m_{\pi^{\pm}}$. 
Figure \ref{S2_1}(a) shows the $M_{00}$
distribution, without any acceptance correction, for the selected events.

Fig.\ref{S2_1}(b) shows the same distribution in the
region close to the threshold $M^2_{00}=4m^2_{\pi^{\pm}}=0.0779 (MeV/c^2)^2$.
The slope change at the threshold is clearly visible. Such an anomaly has not been observed in 
previous experiments.

\section{Interpretation in terms of CHPT}

The sudden change of slope observed in this plot suggests the presence of a threshold "cusp" effect from the decay  
$K^{\pm} \rightarrow \pi^{\pm} \pi^+ \pi^-$ contributing to the  $K^{\pm} \rightarrow \pi^{\pm} \pi^0 \pi^0$ 
amplitude through the charge exchange reaction $\pi^+\pi^- \rightarrow \pi^0\pi^0$. The presence of a cusp 
at $M_{00}^2 = (2\cdot m_{\pi^{\pm}})^2$ in
$\pi^0\pi^0$ elastic scattering due to the effect of virtual $\pi^+\pi^-$ loops has been discussed first by 
Mei\ss ner et al. \cite{Meissner}. 
For the case of
$K^{\pm} \rightarrow \pi^{\pm} \pi^0 \pi^0$ decay Cabibbo \cite{Cabibbo} has proposed a simple re-scattering 
model describing the 
decay amplitude as the sum of two terms representing "unperturbed amplitude" and 
contribution from the $K^{\pm} \rightarrow \pi^{\pm} \pi^+ \pi^-$ decay amplitude 
through $\pi^+\pi^- \rightarrow \pi^0\pi^0$ charge exchange.
The latter contribution is proportional to $a_x$, the S-wave $\pi^+ \pi^-$ charge exchange scattering length.
In the limit of exact isospin symmetry
$a_x =(a_0 - a_2)/3$, where $a_0$ and $a_2$ are the $\pi\pi$ scattering lengths 
in the I=0 and I=2 states, respectively. 

Recently Cabibbo and Isidori \cite{CI} have proposed a more complete model, that takes into account 
all rescattering processes in two-loop approximation. In the limit of exact isospin symmetry the amplitude depends 
on five S-wave scattering lengths, that can be expressed as linear combinations of $a_0$ and $a_2$. 
This model have been used in order to extract $a_0$ and $a_2$ from the fit to the data.
The decay amplitude depends not on $s_3=M^2_{00}$ only, but also on the two $\pi^0\pi^{\pm}$ masses squared 
$s_1,s_2$. We replace them with the average values to avoid multidimensional integration in the fit 
procedure:

\begin{equation}
<s_1>=<s_2>=\frac{3s_0-s_3}{2}
\end{equation}

Where $3s_0=\Sigma s_i = m^2_K + 2m^2_{\pi^0} + m^2_{\pi^+}$.

In order to extract the decay amplitude parameters, the following function $F(M^2_{00})$ 
of the measured $2\pi^0$ mass was fitted to the experimental $M^2_{00}$ spectrum:

\begin{equation}
F=N \int f(a_0,a_2,g_0,h^+,m^2) \Phi(M^2_{00},m^2) dm^2
\end{equation}

Here $f$ is the theoretical distribution from \cite{CI} of kaon decays over the true value of two-pion
invariant mass $m^2$. $\Phi$ is the function, describing the experimental setup resolution and
acceptance, that is obtained by Monte Carlo simulation.

The fit by this model in the interval $0.074 < M_{\pi^0\pi^0}^2 < 0.097 (GeV/c^2)^2$ 
allows to extract four parameters: $(a_0-a_2)m_{\pi^+}$, $a_2m_{\pi^+}$, $g_0$ (the slope),  
$h^+$ (a quadratic term in the Dalitz distribution over $u$ variable)
and one parameter for normalization. It leads to $\chi^2 =154.8$ for 146 degrees of freedom. 
The better fit ($\chi^2 =149.1/145 d.f.$) is obtained by adding to the model 
a term describing the expected formation of  a $\pi^+\pi^-$ atom (``pionium'') decaying 
into $\pi^0\pi^0$ at $M_{00}=2m_{\pi^+}$. 
The fit value for the rate of $K^{\pm}\rightarrow \pi^{\pm} + pionium$ decays 
is $(1.61 \pm 0.66)\cdot10^{-5}$, which is in good
agreement with the predicted value $\sim 0.8\cdot10^{-5}$ \cite{Silagadze}. Sensitivity of the
measured $(a_0-a_2)m_{\pi^+}$ to the pionium production branching ratio is rather small: 
$0.1 \sigma(BR)/BR$.

The model used do not include radiative corrections, which are particularly important near $M_{00}=2m_{\pi+}$, and  
contribute to the formation of the pionium. For this reason a group of seven consecutive bins with the width of 
$0.00015 (GeV/c^2)^2$, centered at $M_{00}=2m_{\pi^+}$ have been excluded from the fit in one of the versions of
analysis. In such a case the measured  $(a_0-a_2)m_{\pi^+}$ value becomes lower by 0.008, that is taken
into account as the contribution to the systematical error.

There were made three independent analyses which use different event selection criteria and different 
Monte Carlo simulations to take into account acceptance and resolution effects. Two of them use a
fast Monte Carlo code, tuned to the experimental resolutions and taking into account the geometry and
kinematics. For the third analysis the full scale GEANT - based \cite{geant} simulation
of the detector is implemented. In all the analyses the extensive checks for the accordance between the simulation 
results and data have been made. The difference between two simulations leads to slightly different second 
derivative of the acceptance as a function of $M_{00}$, that doesn't affect the parameter $(a_0-a_2)m_{\pi^+}$.

Study of systematics shows that no corrections are needed. However, the corresponding uncertainties
were taken into account. For measured $(a_0-a_2)m_{\pi^+}$ parameter the major ones are shown in the
table \ref{system}.

\begin{table*}[htb]
\caption{Contributions to the systematical error.}
\label{system}
\newcommand{\m}{\hphantom{$-$}}
\newcommand{\cc}[1]{\multicolumn{1}{c}{#1}}
\renewcommand{\tabcolsep}{2pc} % enlarge column spacing
\renewcommand{\arraystretch}{1.2} % enlarge line spacing
\begin{tabular}{@{}ll} \hline
Excluding of pionium region from the fit & $0.008$ \\
Cuts variation                           & $0.004$ \\
Z - dependence                           & $0.009$ \\
MC-related  difference                   & $0.006$ \\ \hline 
TOTAL (adding in quadrature)             & $0.014$ \\ \hline 
\end{tabular}\\[2pt]
\end{table*}

\section{Preliminary result and discussion}

A critical parameter of the models \cite{Cabibbo} and \cite{CI} is the ratio $R= A_{++-}/A_{+00}$ between
weak amplitudes of $K^{\pm} \rightarrow \pi^{\pm} \pi^+ \pi^-$ and $K^{\pm} \rightarrow \pi^{\pm} \pi^0 \pi^0$ 
decays.
The corresponding "external" uncertainty should be complemented by an additional theoretical error of $\pm 5\%$
as the result of neglecting higher-order terms and radiative corrections. 

Taking into account these uncertainties the following preliminary result is obtained:

\begin{eqnarray}
\nonumber
(a_0-a_2)m_{\pi^+} = 0.281 \pm 0.007 (stat) \\ 
\nonumber
                           \pm 0.014(syst) \pm 0.014(extern)
\end{eqnarray}

This value is compatible withing the errors with the results obtained in the E865 BNL \cite{BNL}
and the Dirac \cite{Dirac} experiments, as well as with the CHPT calculation 
result \cite{CHPT1}, \cite{CHPT2} ($(a_0-a_2)m_{\pi^+} = 0.265 \pm 0.004$). It is also 
in good agreement with the result of fit to low energy experimantal data using 
dispersion relations \cite{Yndurain} ($(a_0-a_2)m_{\pi^+}=0.278 \pm 0.016$). 

In the forthcoming stages of analysis we plan to decrease the systematical uncertainty and to take
into account the isospin symmetry breaking.
At tree level, omitting the one-photon exchange diagrams, isospin symmetry breaking contributions
to the elastic $\pi\pi$ scattering amplitude can be expressed as a function of one parameter
$\epsilon = (m_{\pi^+}-m_{\pi^0}^2)/m_{\pi^+}^2 = 0.065$ \cite{breaking}. In particular, the
amplitudes at the threshold of $2\pi^{\pm}$ mass are corrected by a factor of $(1-\epsilon)$ for
the processes $\pi^+\pi^+ \rightarrow \pi^+\pi^+$, $\pi^+\pi^0 \rightarrow \pi^+\pi^0$ and
$\pi^0\pi^0 \rightarrow \pi^0\pi^0$; by $(1+\epsilon)$ for  $\pi^+\pi^- \rightarrow \pi^+\pi^-$;
and by $(1+\epsilon/3)$ for  $\pi^+\pi^- \rightarrow \pi^0\pi^0$. Implementing of such a correction
will lead to the corresponding decrease of measured $(a_0-a_2)m_{\pi^+}$.

\end{document}